# Directional self-locomotion of active droplets enabled by nematic environment

Mojtaba Rajabi [1,†], Hend Baza [1,†], Taras Turiv [2], Oleg D. Lavrentovich [1,2,*]

[1]Department of Physics, Kent State University, Kent, OH 44242, USA

[2]Advanced Materials and Liquid Crystal Institute, Chemical Physics Interdisciplinary Program, Kent State University, Kent, OH 44242, USA

[*] Corresponding author. Email: olavrent@kent.edu

[†] These authors contributed equally to this work

**Abstract.** Active matter comprised of self-propelled interacting units holds a major promise for extraction of useful work from its seemingly chaotic out-of-equilibrium dynamics. Streamlining active matter to produce work is especially important at microscale, where the viscous forces prevail over inertia and the useful modes of transport require very specific non-reciprocal type of motion. Here we report that microscopic active droplets representing aqueous dispersions of swimming bacteria *Bacillus subtilis* show a unidirectional propulsion when placed in an inactive nematic medium. Random motion of bacteria inside the droplet is rectified into a directional self-locomotion of the droplet by the polar director structure that the droplet itself creates in the surrounding nematic through anisotropic molecular interactions at its surface. Droplets without swimming bacteria show no net displacement. The trajectory of the active droplet can be predesigned as rectilinear or curvilinear by patterning the molecular orientation of the nematic medium. The effect demonstrates that swimming at microscale can be achieved at the expense of broken spatial symmetry of the medium; it can be used in development of micromachines.

Active droplets are spatially restricted fluid volumes containing self-propelled and interacting units. They are gaining an increasing interest as a model system [1-10] to understand dynamics of microscopic biological organisms such as cells and bacteria that developed sophisticated modes of propulsion and self-replication. Self-propulsion at microscales is difficult since viscosity prevails over inertia, and the moving microorganisms should use non-reciprocal modes of movement, such as rotation of a helicoidal flagellum [11]. Another challenge is maintaining a certain direction of propulsion, since Brownian motion randomizes trajectories of microparticles [12]. Most living and synthetic microswimmers demonstrate an "enhanced" version of the Brownian motion. At time scales shorter than the time of the particle realignment by orientational diffusion, the motion is directional, but at longer time scales, the motion becomes diffusive with zero net displacement, albeit with a diffusion coefficient that is substantially higher than the one of the "passive" Brownian particle [12]. Sanchez et al.[2] reported such an enhanced Brownian motion of active disk-like droplets of kinesin-activated dispersions of microtubules, in which chaotic dynamics of microtubules causes translations of the droplet because of its frictional contact with the substrate; at long time scales, rotational diffusion makes the motion unbiased and Brownian-like. A similar enhanced Brownian dynamics is observed for *E. coli*-containing water droplets dispersed in an isotropic oil environment [10]. As stressed by Marchetti [13], an important challenge is to find the way to control the propulsion direction of active droplets. A controllable propulsion would allow one to extract a useful work at microscale, which is a major promise of active matter in terms of practical applications. Extraction of work from active matter has been achieved for ratchet gears that rotate when placed in an active bacterial bath [14,15]; in contrast, a controllable rectilinear or curvilinear propulsion remains elusive.

In this work, we demonstrate controllable rectilinear and curvilinear propulsion of active droplets enabled by spatially broken symmetry of the local environment that the active droplet creates when placed in an orientationally ordered nematic fluid. The active droplets represent aqueous dispersions of motile bacteria *Bacillus subtilis*. These droplets are dispersed in a passive hydrophobic thermotropic nematic. Molecular orientation of the nematic is perpendicular at the surface of the active droplet, producing a polar fore-aft asymmetric environment around the droplet. Active bacterial flows inside the droplet trigger flows in the nematic surrounding that are rectified by the polar symmetry of the latter to yield a directional locomotion of the droplet. In absence of active bacteria, the droplets show only Brownian diffusion with no net displacement. The rectilinear and curvilinear trajectories of active droplets are predesigned by patterning the orientation of the nematic medium.

We use *B. subtilis* strain 1085, which is a rod-shaped bacterium 5–7 µm long and ~0.7 µm in diameter. Two types of active droplets are prepared. One is based on an isotropic aqueous Terrific Broth (TB) (Sigma Aldrich) as a dispersion medium. The second medium for bacteria is a lyotropic chromonic liquid crystal (LCLC), 13 wt.% dispersion of disodium cromoglycate (DSCG) (Alfa Aesar) in TB. The active droplets are emulsified in the thermotropic nematic LC MAT-03-382 (Merck) of low birefringence (to facilitate optical microscopy observations). Both isotropic and LCLC-based active droplets contained a very small amount ($\leq 0.05$ wt%) of lecithin (Sigma Aldrich) which sets a perpendicular orientation of the nematic director at the interface. Our LCLC-based system is an inverted active analog of the nematic-in-LCLC emulsion, in which the thermotropic nematic droplets are dispersed in a LCLC with an added surfactant.[16]. The emulsion in Ref.[16] does not contain any swimmers, but nematic droplets can propel thanks to the Marangoni effect. The emulsions of active droplets in the nematic are filled into cells of thickness $d = 200$ µm, formed by two parallel glass plates. The plates are coated with polyimide PI2555 (Merck) layers and buffed to achieve a unidirectional planar alignment, $\hat{\mathbf{n}}_0 = (1,0,0)$ in Cartesian coordinates.

The surfactant lecithin molecules diffuse to the water-nematic interface and impose perpendicular alignment of the nematic director at the surface of active droplets. To match the overall uniform alignment of the director imposed by the bounding glass plates, each droplet acquires a satellite topological defect: either a point defect, the so-called hyperbolic hedgehog[17], Fig. 1a-c, or an equatorial disclination ring[18-20], Fig. 2a,b. We call these the hedgehog (H) and Saturn ring (SR) configurations, respectively. The H-structure typically forms around smaller droplets, of diameter $2R \leq 140$ µm and the SR forms around larger droplets, that almost fill the entire 200 µm space between the glass plates: the SR is stabilized by the unidirectional anchoring at the bounding plates[19,21]. The H and SR configurations differ dramatically in their symmetry and in impact on the dynamics of active droplets.

***The hedgehog structure*** is of a dipolar symmetry, Fig. 1a-c, and rectifies the chaotic motion of bacteria inside the droplets into a polar locomotion of the droplet along the *x*-axis, Fig. 1d-f. Both the LCLC-based, Fig. 1d, and isotropic active droplets, Fig. 1e, with an accompanying hedgehog propel themselves along $\hat{\mathbf{n}}_0 = (1,0,0)$, with the hedgehog leading the way, $\mathbf{v} = (v,0,0)$, Fig. 1b and Supplementary Movie 1. This self-produced locomotion is not strictly unidirectional, as the droplet makes steps forward and backward along $\hat{\mathbf{n}}_0 = (1,0,0)$, as illustrated by a histogram in Fig. 1f for displacements $\Delta x$ measured over time intervals of 0.05 s; $\Delta x > 0$ for the displacement towards the hedgehog and $\Delta x < 0$ for the displacement away from the hedgehog. As easy to see in Fig.1f, the motion with $\Delta x > 0$ prevails. Displacements of the active droplets along the *y*-axis perpendicular to $\hat{\mathbf{n}}_0$ are of an "enhanced" Brownian type, producing no net displacement over time, Fig. 1d,e. The speed of active droplets along $\hat{\mathbf{n}}_0$ depends on the interior composition: LCLC-based droplets move about 10 times faster than their isotropic counterparts. For example, an isotropic droplet of a diameter $2R = 90$ μm and bacterial concertation $c = 20 c_0$ (here $c_0 = 8 \times 10^{14}$ cells/m$^3$ is the concentration of bacteria at the end of their exponential growth in an incubator) moves with an average speed of 2 μm/min, Fig. 1e, while an LCLC active droplet with the same parameters moves more than an order of magnitude faster, Fig. 1d. The propulsion speed of active droplets increases with their diameter, Fig. 1g, and with the concentration of bacteria, Fig. 1h. As the time goes by and the bacteria consume nutrients, their activity diminishes and the speed of propagation decreases, Fig. 1i; droplets that do not contain bacteria show only Brownian diffusion and no locomotion.

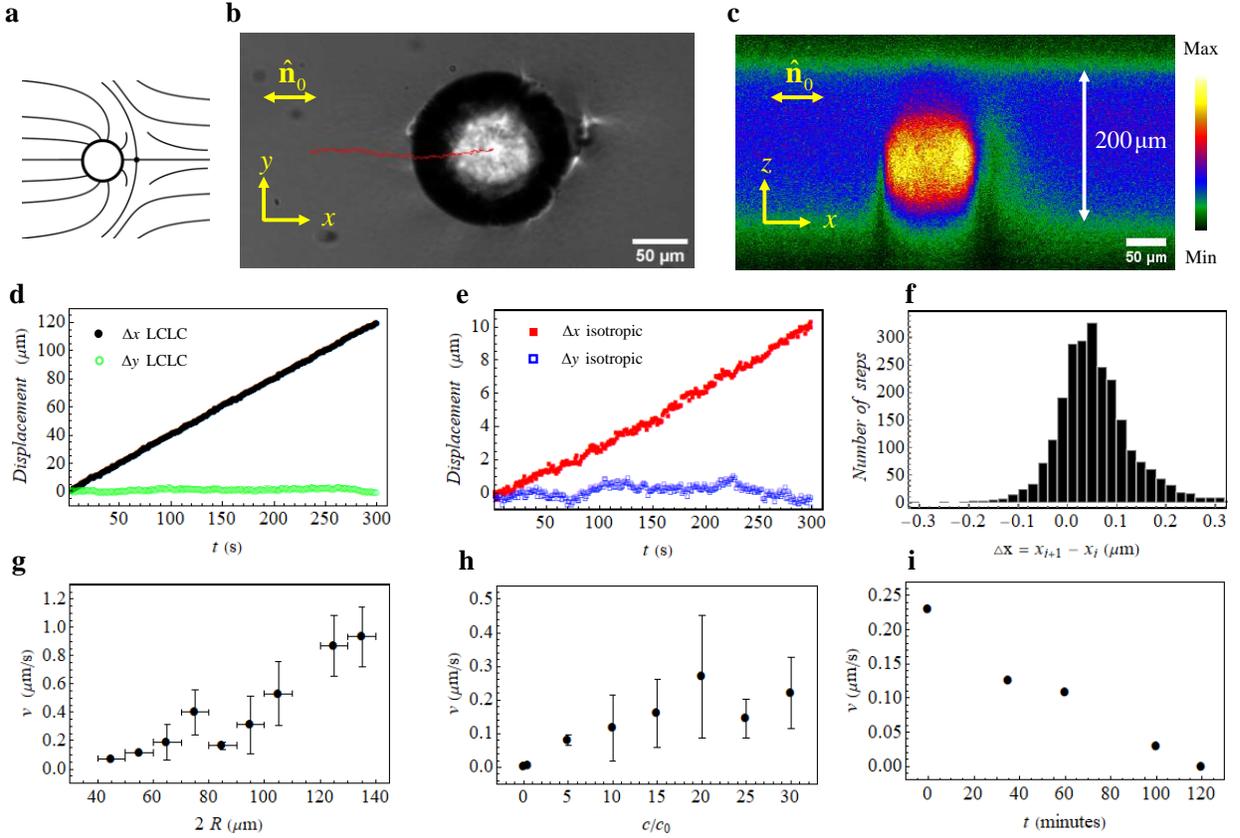

**Fig. 1. Self-propelled active H-droplets. a**, director configuration around a sphere with perpendicular surface anchoring that produces a point defect-hyperbolic hedgehog; **b**, optical microscopy texture (no polarizers) of a self-propelled droplet; the red trajectory is traced during 10 min; **c**, fluorescent confocal polarizing microscopy texture of the vertical cross-section of a sample with the LCLC-based active droplet in the middle; color corresponds to the intensity of fluorescent signal from the sample; **d**, displacements $\Delta x$ along $\hat{\mathbf{n}}_0$ and $\Delta y$ in the perpendicular directon, measured every 1 s for LCLC-based active droplet, $2R = 90 \, \mu m$, $c = 20 c_0$; **e**, the same for an isotropic active droplet with $2R = 90 \, \mu m$, $c = 20 c_0$; **f**, histogram of droplet displacements measured within 0.05 s intervals along the $x$-axis, positive steps are towards the hedgehog; $2R = 135 \, \mu m$, $c = 20 c_0$; propulsion speed $v$ vs (**g**) droplet diameter, $c = 20 c_0$; (**h**) bacterial concentration, $2R = 90 \, \mu m$; (**i**) time since the start of observations; $2R = 120 \, \mu m$, $c = 5 c_0$. All data, except for part (**i**), are obtained within 30 min since sample preparation.

***The Saturn ring*** configuration around active droplets is of a quadrupolar symmetry, Fig. 2a,b and shows no rectified propulsion. The active droplets with SR engage in anisotropic anomalous diffusion, Fig. 2c,d, typical of colloidal dynamics in a passive nematic environment [22,23]; with the displacements along the director on average longer than in the perpendicular direction. This dynamics yields no net displacement when averaged over 1-10 min. Both SR and H configurations fluctuate in time, being perturbed by conventional factors characteristic of passive systems, such as the fluctuations of the director, and by active flows produced by swimming bacteria inside the droplets, similarly to the case described by Guillamat et al [20] for SR structures around active spherical shells. Activity-induced fluctuations of SR around droplets of an intermediate size, such as the one in Fig. 2e with $2R = 130$ μm, result in dramatic shifts of the SR from the equatorial location, which we quantify by the ratio $r/R$, where $r$ is the distance of the SR center from the droplet center, Fig. 2f. These SR shifts change the director field symmetry from quadrupolar to dipolar; the SR can even collapse into the H structure (in which case $r/R > 1$). The propulsion velocity of the active droplets is proportional to the asymmetry ratio $r/R$, Fig. 2e,f. When the SR shifts to the left, the droplet propels to the left as well; when the SR shifts to the right, so does the droplet; the highest propulsive speed is achieved when the SR shrinks into the hedgehog H, thus maximizing $|r/R|$, Fig. 2 e,g and Supplementary Movie 2.

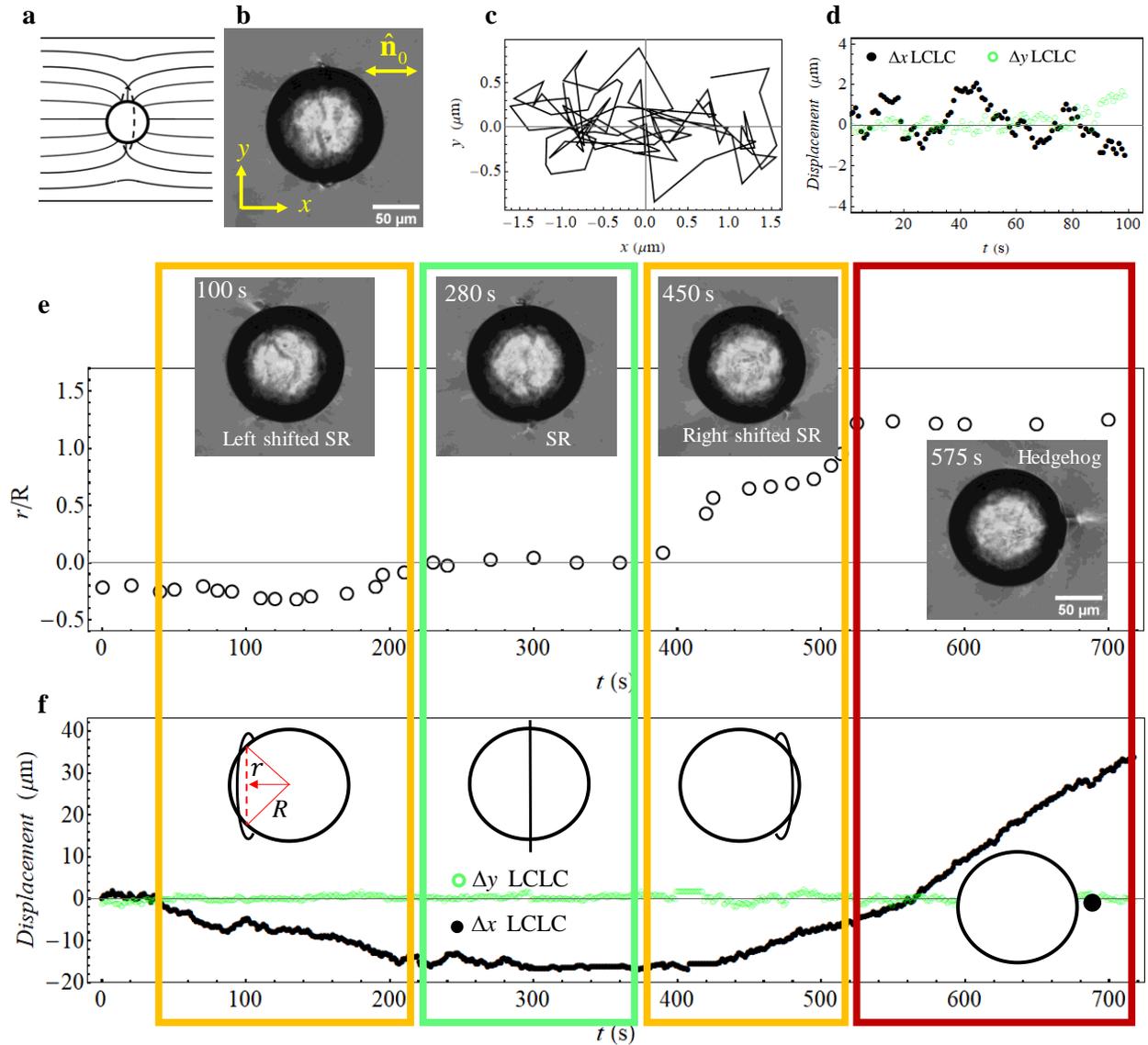

**Fig. 2. Active droplet accompanied by a Saturn ring. a**, SR director configuration around a sphere with perpendicular surface anchoring; **b**, optical microscopy texture (no polarizers) of a SR droplet of quadrupolar symmetry; **c**, trajectory of the SR droplet traced during 100 s; no net displacement; **d**, displacements $\Delta x$ along $\hat{\mathbf{n}}_0$ and $\Delta y$ in a perpendicular directon, measured every 1 s; **e**, SR around the droplet is strongly shifted by active flows, first to the left ($t < 220\,\mathrm{s}$), then back to the equatorial position ($220\,\mathrm{s} < t < 370\,\mathrm{s}$), then to the right ($370\,\mathrm{s} < t < 520\,\mathrm{s}$); SR eventually shrinks into a hedgehog ($t > 520\,\mathrm{s}$); the plot shows time evolution of the asymmetry degree $r/R$; **f**, $(x, y)$ coordinates of the same active droplet as in (**e**) within the same time interval;

velocity of self-propulsion grows with the degree of asymmetry $|r/R|$. All data are obtained within 30 min since sample preparation for the same LCLC-based droplet with $2R = 130\,\mu m$, $c = 20\,c_0$.

*Mechanism of self-locomotion.* As established above, self-locomotion is pertinent only to the active droplets that develop the H-structures of the nematic director around them; the SR-droplets are not motile. The reason is the dipolar symmetry of the director around the H-droplets and its ability to rectify the random flows produced by the bacterial swimmers inside the droplets.

Motile bacteria at elevated concentrations such as explored in this work, produce turbulent flows in both aqueous [24] and LCLC environments [25] with vortices of a typical size $(10-40)\,\mu m$. Particle image velocimetry (PIV) tracking of the fluorescent tracers reveals that the interior of the active droplets exhibits similar chaotic flows with numerous vortices that are poorly correlated in space and time. The instantaneous maps of velocity and color-coded vorticity $\omega = \partial v_y/\partial x - \partial v_x/\partial y$ in Fig.3a show that the spatial correlations do not extend over distances larger than $(30-40)\,\mu m$, while time correlations are shorter than 5-10 s. Averaging active flow patterns over few minutes diminishes the overall velocities and vorticities; the only regular feature is some alignment of flows along the water-nematic interface, Fig. 3b.

The active flows inside the droplet transfer through the water-nematic interface and trigger flows of the nematic, Fig. 3c. Unlike the interior flows, the exterior flows are rectified over time by the dipolar director H-structures and produce a net flow, Fig.3d. In the droplet's coordinate system, the nematic flows from the hedgehog towards the droplet, Fig.3d. In the laboratory system, it means that the active droplet self-propels with the hedgehog leading the way. Rectification of the flow by the polar director structure is rooted in the well-known coupling of the flow and the director field of nematics [26]. Note here that the propulsive ability of the active droplets cannot be explained by the Marangoni effect associated with a dynamic redistribution of lecithin and gradients of the interfacial tension, as droplets with the surfactant but no active bacteria show no rectified motility, even when the concentration of lecithin is increased to 1wt%. Droplets of a diameter smaller than $30\,\mu m$ do not show rectified flows and thus do not exhibit self-locomotion, Fig.1g, apparently because the number of bacteria is not sufficient to produce strong flows. Another factor diminishing self-locomotion ability of small droplets might be the decrease of surface anchoring ($\sim R^2$) as compared to bulk elasticity of the nematic ($\sim R$). If the droplet is large, with the SR structure of quadrupolar symmetry, the time-averaged flows in the nematic are left-right and fore-aft symmetric, producing no net displacement.

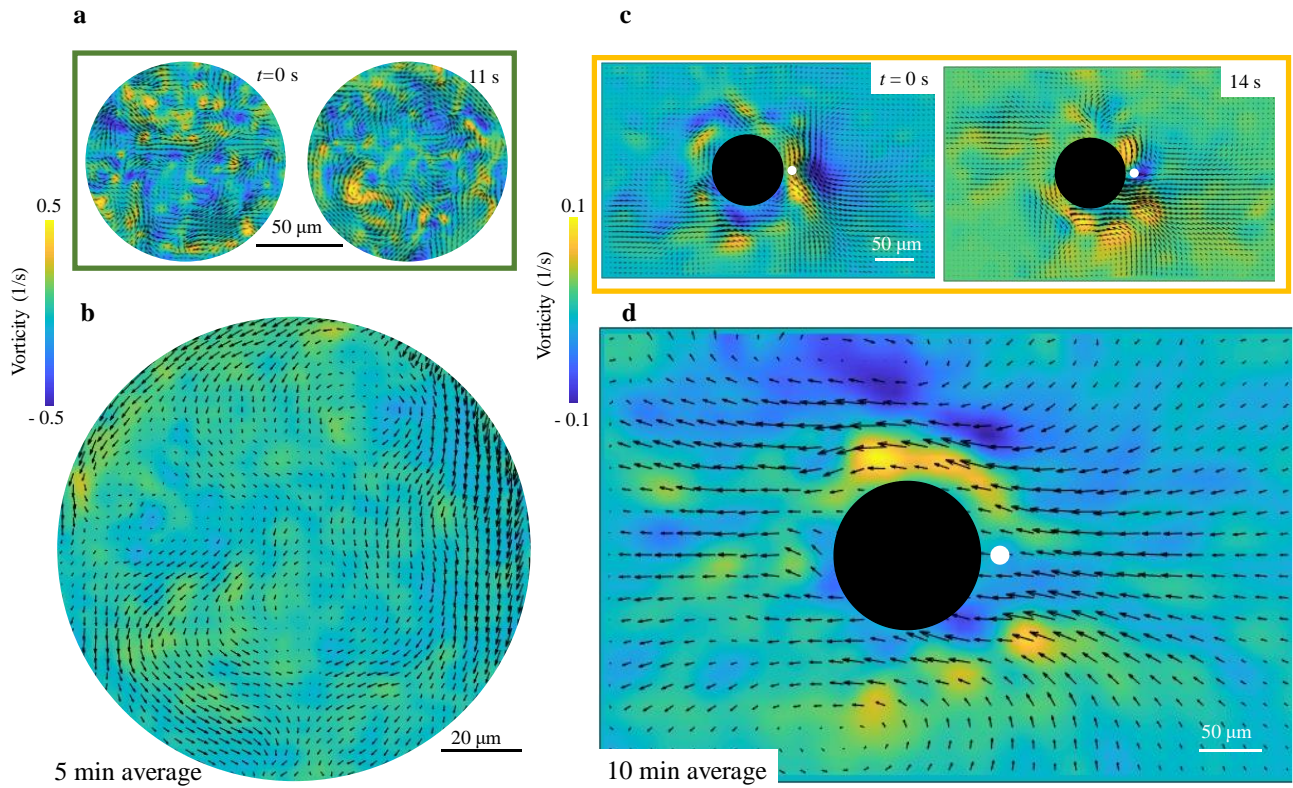

**Fig. 3. Activity triggered flows (a,b) inside and (c,d) outside the H-droplet. a**, instantaneous velocity (the maximum length of velocity vectors corresponds to $6\,\mu m/s$) and color-coded vorticity of active flows within an active droplet with $2R = 125\,\mu m$, $c = 20\,c_0$; **b**, the same droplet, velocity and vorticity are averaged over 5 min; maximum velocity is 1 $\mu m/s$; **c**, velocity (maximum 2 $\mu m/s$) and vorticity of active flows in the nematic outside an active droplet with $2R = 110\,\mu m$, $c = 20\,c_0$; **d**, the same droplet, velocity (maximum 0.2 $\mu m/s$) and vorticity are averaged over 10 min. All data are obtained for LCLC-based droplets within 30 min since sample preparation.

*Curvilinear trajectories.* So far, we described cells with a uniform director of the thermotropic nematic. The director can also be pre-aligned in various spatially-varying $\hat{\mathbf{n}}_0(x, y)$ patterns by plasmonic photoalignment technique [27]. These patterns command the droplets to follow a path set by $\hat{\mathbf{n}}_0(x, y)$, provided the director does not change its orientation too strongly over the length scales $\sim 2R$ or smaller. When a reasonably small droplet ($30\,\mu\text{m} < 2R < 140\,\mu\text{m}$ in a cell of thickness $200\,\mu\text{m}$) is dispersed in the nematic with such a smoothly-changing $\hat{\mathbf{n}}_0(x, y)$, it also develops an H-structure and self-locomotion ability. Both the time-averaged structural vector connecting the center of the droplet to the hedgehog and the velocity vector of the self-propelled droplet are parallel to the local $\hat{\mathbf{n}}_0$, realigning themselves as $\hat{\mathbf{n}}_0$ reorients. As an example, we design a circular concentric director field, Fig.4, by photopatterning the substrates. The active droplets dispersed in such a patterned environment move along circular trajectories, as dictated by the patterned director $\hat{\mathbf{n}}_0$, Fig. 4 and Supplementary Movie 3.

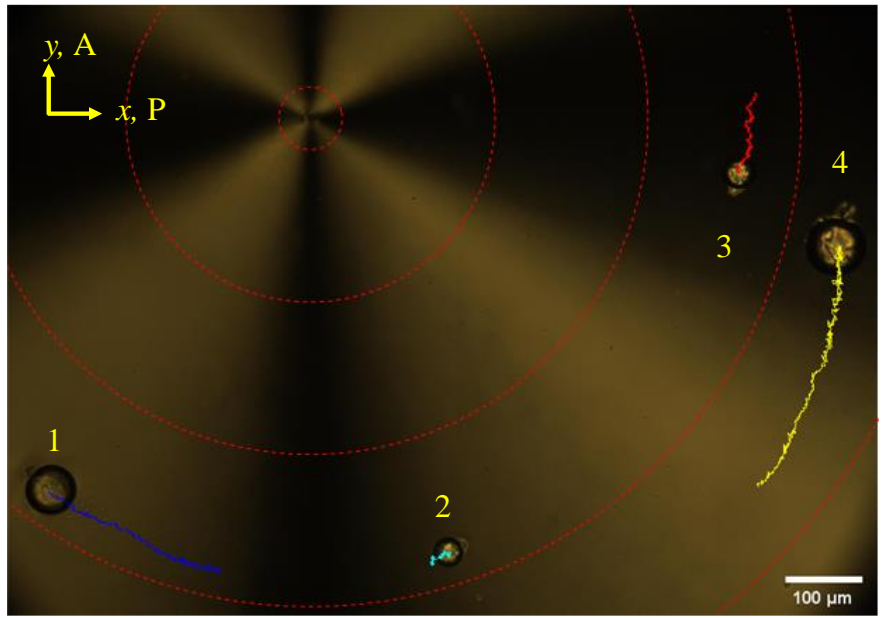

**Fig. 4. Circular trajectories of active H-droplets in a nematic with a circular prepatterned director (dashed lines).** Droplets of diameter $2R = 68\,\mu\text{m}$ (1), $40\,\mu\text{m}$ (2), $50\,\mu\text{m}$ (3) and $85\,\mu\text{m}$ (4). All data are obtained for LCLC-based droplets with bacterial concentration $c = 5c_0$, within 40 min since sample preparation. Polarizing optical microscopy with crossed polarizers A and P.

**Conclusions.** We demonstrated a novel way to exteract work from random motion of microswimmers and to power rectilinear and curvilinear self-locomotion of active droplets. The concrete example is based on swimming bacteria, dispersed in aqueous media with nutriens and with or without additional ingredients such as the lyotropic liquid crystal. These dispersions are emulsified as active droplets in a thermotropic nematic with an added surfactant that produces a locally radial orientation of the nematic director around the droplet. This radial local director matches the uniform far field through topological defects, either a "Saturn ring" of a disclination loop, or a point defect-hedgehog. The hedgehog structure is of a polar symmetry. When the active droplet develops the fore-aft asymmetric hedgehog structure, it acquires the ability to propel itself, along a rectilinear or curvilinear trajectory that is set by the overall director of the nematic.

The active flows generated by bacteria inside the droplet trigger flows in the nematic that are rectified by the polar director structure into a net directional flow. The efficiency of self-propulsion is directly related to the degree of polar asymmetry, as shown by the example with a strongly fluctuating Saturn ring in Fig. 2 and in Supplementary Movie 2. The droplets formed by dispersions of bacteria in lyotrtopic chromonic liquid crystals show a faster speed than their isotropic counterparts, Fig. 1d,e; the enhancement is related to a higher viscosity of the LCLC material and thus more efficient momentum transfer through the interface. The droplets with hedgehogs move faster when they are larger and when the concentration of bacteria is higher. Droplets without motile bacteria do not propel. By patterning the overall director field, one can predesign the geometry of trajectories. The demonstrated mechanism adds to the set of principles of microscale swimming described by Purcell [11], as the motile object is essentially a sphere with a chaotic internal motion and yet it can self-propel by changing the symmetry of its immediate passive fluid environment.

The demonstrated rectilinear and curvilinear self-locomotion of active droplets enabled by a liquid crystal environment can find applications in development of micromachines. In the particular example presented in this work, the self-locomotion is powered by swimming bacteria. The demonstrated principle should be extendable to other systems, for example, systems that use synthetic microswimmers. In the latter case, the active droplets might not be necessarily water based; one can envision an inverted system with active droplets based on oils, dispersed in a lyotropic liquid crystal medium.

**Methods.** The bacteria *B. subtilis* are initially grown on Lysogeny broth (Miller composition from Teknova, Inc.) agar plates at 35°C for 12-24 hrs, then a colony is transferred to a Terrific Broth (TB) liquid medium and grown in shaking incubator at temperature 35°C for 7-9 hrs. Sealed vials are used to grow the bacteria to increase resistance to oxygen starvation. The bacteria concentration in the growth medium was monitored by measuring optical density. The bacteria are removed from incubator at the end of their exponential growth stage, at the concentration $c_0 = 8 \times 10^{14}$ cell/m$^3$, and extracted from the liquid medium by centrifugation. They are then dispersed in TB or in TB with DSCG. The bacteria-containing aqueous dispersion is mixed with the thermotropic nematic in volume proportion 1:50 and vortexed to achieve the emulsion. A cell formed by two glass plates is filled with the active emulsion by capillary forces. The plates are coated with polyimide PI2555 (Merck) layers, buffed to achieve a unidirectional planar alignment. Circularly patterned cells are aligned by the plasmonic metamask photoalignment technique [27].

The location of the droplet along the *z*-axis normal to the bounding plates is determined by fluorescent confocal polarizing microscopy [28]. The nematic is doped with the fluorescent hydrophobic dye N,N′-Bis(2,5-di-tert-butylphenyl)-3,4,9,10-perylenedicarboximide (BTBP) at 0.0025 wt%, and the active droplets are doped with a hydrophilic dye Acridine Orange at 0.006 wt%. The in-plane motion of active droplets is observed under a Nikon TE2000 optical microscope equipped with a camera CMOS (Emergent HS-20000C); the trajectories are tracked using ImageJ software [29]. The flows inside and outside of the droplet are visualized by tracking fluorescent polystyrene Suncoast Yellow spheres of diameter 200 nm (Bangs Lab), excitation wavelength 540 nm and emission wavelength 600 nm. These wavelengths are safe for bacteria and do not alter their activity over the time of experiments. The fluorescent spheres are dispersed in both media and tracked by particle image velocimetry (PIV) [30].

**Acknowledgments**

**Funding:** The work is supported by NSF grants DMR-1905053 (analysis of dynamics), CMMI-1663394 (preparation of plasmonic metamasks for patterned cells), DMS-1729509 (preparation of bacterial dispersions), and by Office of Sciences, DOE, grant DE-SC0019105 (development of the alignment layers).

**Author contributions:** M.R. and H.B. performed the experiments, M.R., H.B. and T.T. analyzed the data, O.D.L. conceived and supervised the project, M.R. and O.D.L. wrote the manuscript with the input from all co-authors.

**Competing interests:** The authors declare no competing interests.

**Data and materials availability:** All data in support of the reported findings are available from the corresponding author upon reasonable request.